\documentclass[sigconf,nonacm]{acmart}

\usepackage{csquotes}
\usepackage[english]{babel}
\usepackage[utf8]{inputenc}
\usepackage{hyphenat}
\usepackage{algorithm}
\usepackage{algpseudocode}
\usepackage{enumitem}
\usepackage[dvipsnames]{xcolor}
\usepackage{tcolorbox}
\usepackage{listings}
\usepackage{adjustbox}
\usepackage{zi4}
\usepackage{stfloats}
\usepackage{balance}
\usepackage{multirow}
\lstdefinestyle{coverage}{
  basicstyle=\ttfamily\footnotesize,
  frame=single,
  breaklines=true,
  columns=fullflexible
}
\usepackage{balance}
\usepackage{multicol}
\usepackage[bottom,flushmargin]{footmisc}
\newcommand{\zw}{\fontfamily{pcr}\selectfont}

\newcommand{\pq}{\textsc{PrediQL}}

\tcbuselibrary{listings,breakable}
\usepackage{booktabs,siunitx,threeparttable}
\makeatletter
\def\@ACM@checkaffil{
    \if@ACM@instpresent\else
    \ClassWarningNoLine{\@classname}{No institution present for an affiliation}%
    \fi
    \if@ACM@citypresent\else
    \ClassWarningNoLine{\@classname}{No city present for an affiliation}%
    \fi
    \if@ACM@countrypresent\else
        \ClassWarningNoLine{\@classname}{No country present for an affiliation}%
    \fi
}
\makeatother

\newcommand{\np}{\fontfamily{lmtt}\selectfont}

\definecolor{commentcolor}{RGB}{63, 127, 95}
\definecolor{functioncolor}{RGB}{0, 0, 128}
\definecolor{keywordcolor}{RGB}{128, 0, 0}
\definecolor{builtin}{RGB}{51, 51, 255}

\newtcolorbox{rqbox}[1][]{
  colback=gray!5!white,
  colframe=gray!50!black,
  fonttitle=\bfseries,
  title=#1,
  title after break={},
  boxrule=0.5pt,
  arc=2pt,
  left=6pt,
  right=6pt,
  top=4pt,
  bottom=4pt
}

\title{\pq: Automated Testing of GraphQL APIs with LLMs}

\author{
Shaolun Liu$^{1}$\textsuperscript{*},
Sina Marefat$^{2}$\textsuperscript{*},
Omar Tsai$^{1}$, Yu Chen$^{1}$,\\ 
Zecheng Deng$^{1}$,
Jia Wang$^{1}$,
Mohammad A. Tayebi$^{1}$}
\affiliation{%
  \institution{$^{1}$ Simon Fraser University, Canada}
}
\affiliation{%
  \institution{$^{2}$ K. N. Toosi University of Technology, Iran}
}
\email{shaolun.liu@sfu.ca, sina.marefat@email.kntu.ac.ir, omar@ztasecurity.com,}
\email{ yca518@sfu.ca, zda35@sfu.ca, jwa454@sfu.ca, tayebi@sfu.ca}

\begin{document}

\sloppy

\begin{abstract}
GraphQL’s flexible query model and nested data dependencies expose APIs to complex, context-dependent vulnerabilities that are difficult to uncover using conventional testing tools. Existing fuzzers either rely on random payload generation or rigid mutation heuristics, failing to adapt to the dynamic structures of GraphQL schemas and responses. We present \pq, the first retrieval-augmented, LLM-guided fuzzer for GraphQL APIs. \pq~combines large language model reasoning with adaptive feedback loops to generate semantically valid and diverse queries. It models the choice of fuzzing strategy as a multi-armed bandit problem, balancing exploration of new query structures with exploitation of past successes. To enhance efficiency, \pq~retrieves and reuses execution traces, schema fragments, and prior errors, enabling self-correction and progressive learning across test iterations. Beyond input generation, \pq~integrates a context-aware vulnerability detector that uses LLM reasoning to analyze responses, interpreting data values, error messages, and status codes to identify issues such as injection flaws, access-control bypasses, and information disclosure. Our evaluation across open-source and benchmark GraphQL APIs shows that \pq~achieves significantly higher coverage and vulnerability discovery rates compared to state-of-the-art baselines. These results demonstrate that combining retrieval-augmented reasoning with adaptive fuzzing can transform API security testing from reactive enumeration to intelligent exploration.
\end{abstract}

\keywords{GraphQL Security, API Fuzzing, Large Language Models, Retrieval-Augmented Generation, Multi-Armed Bandit Learning}

\maketitle

\footnotetext[1]{Equal contribution.}

\section{INTRODUCTION}
Modern software systems are often built from many independent microservices that communicate through well-defined APIs. Among contemporary API styles, GraphQL has gained significant adoption because it allows clients to request precisely the data they need through a single flexible endpoint. This design improves efficiency and mitigates the problem of over-fetching, common in REST~\cite{aws_graphql_rest_2024,andersson2021rest_graphql_grpc} or gRPC~\cite{aws_graphql_rest_2024} APIs.

According to a 2024 industry survey, approximately 61\% of respondents reported using GraphQL in production, and about 10\,\% indicated that they were replacing REST with GraphQL in their systems~\cite{techtarget2024graphql,hygraph2024survey}. However, this widespread adoption has also exposed security weaknesses in many deployments. A recent study reported that roughly 69\,\% of the scanned public GraphQL API services suffered from unrestricted resource consumption vulnerabilities, making them susceptible to denial of service (DoS) attacks via deep-nested or costly queries~\cite{escape2024security,ibm_dos_graphql_2023,fastly_graphql_security_2022}. These issues highlight that GraphQL introduces unique security challenges, such as unbounded query depth, schema exposure, injection in nested arguments, and inconsistent access control across linked queries and mutations~\cite{graphql_security_page_2025}. Consequently, effective testing of GraphQL APIs requires more than traditional input fuzzing; it must incorporate a contextual understanding of schema relationships and multistep interactions.  

Existing testing tools fall into two main groups: black-box fuzzers and schema-aware fuzzers. Black-box fuzzers explore input space randomly, but do not respect GraphQL structure. Schema-aware tools, such as  \textsc{EvoMaster}~\cite{belhadi2023evomaster,arcuri2021evomaster,belhadi2023graphql}, improve test coverage by using the schema; but rely on simple template mutations and ignore cross-field or cross-operation dependencies. Recent systems such as \textsc{GraphQLer}~\cite{graphqler2025} take an important step by analyzing producer–consumer relationships between queries and mutations, discovering vulnerabilities missed by older tools. However, even these approaches remain static; their generation logic does not adapt to execution feedback or changing context. In short, current GraphQL fuzzers cannot yet reason adaptively about API behavior.

Recent advances in large language models (LLMs) offer a new direction on how to build intelligent, feedback-driven fuzzers. LLMs can reason about structured input formats, infer valid parameters from schema fragments, and generate meaningful queries~\cite{ouyang2022instructgpt,madaan2023selfrefine}. When combined with retrieval mechanisms that use past traces, schema segments, or error logs~\cite{lewis2020rag,reimers2019sbert,johnson2019faiss}, the fuzzer can refine future inputs based on what it has already learned. Recent LLM-assisted fuzzing solutions~\cite{xia2023fuzz4all,fuzzgpt2024,deng2022titanfuzz,meng2024hyllfuzz,zhang2025dfuzz,surveyllmfuzz2024} show that generative reasoning improves coverage in binary and protocol fuzzing. Yet, no prior research has applied LLM-based fuzzing to GraphQL, whose nested queries and dependency-rich structure require contextual reasoning and adaptive feedback. Work such as WENDIGO~\cite{ieee10579536} targets denial-of-service discovery in GraphQL using deep reinforcement learning, and Perera et~al.~\cite{perera2025enhancing} explore detecting malicious GraphQL queries with LLMs and neural classifiers, but neither focuses on adaptive fuzzing.

To close this gap, we present \pq, an automated GraphQL testing framework that joins schema introspection, retrieval-augmented LLM prompting~\cite{lewis2020rag}, multi-armed bandit learning~\cite{lattimore2020bandit,agrawal2011ts,cavenaghi2021nonstationary} and self-correction into a single feedback loop. \pq~treats the LLM as a guided component rather than an oracle. It first extracts schema information through introspection and then retrieves relevant examples and past errors to generate the query in real feedback. A bandit-based selector dynamically chooses between different prompting strategies, balancing exploration and exploitation. This design allows \pq~to generate valid, diverse, and context-sensitive queries that explore deeper parts of the schema and reveal complex vulnerabilities. 

We evaluated \pq~using multiple large language models across a diverse set of open-source and benchmark GraphQL APIs. The results show that \pq~consistently improves test coverage, by an average of 16\% with a maximum improvement of 50\%, and discovers more context-dependent vulnerabilities compared to established baselines. Among the tested configurations, the {\zw GPT-5 Mini} achieved the highest coverage, while smaller models like Llama-3-8B offered competitive results with reduced computational cost. In addition to broader coverage, \pq~demonstrates stronger vulnerability detection capabilities, accurately identifying injection flaws, access control bypasses, and information disclosure cases that existing tools often miss.

In summary, this paper makes three main contributions: (1) We present PrediQL, the first retrieval-augmented, LLM-guided GraphQL fuzzer with adaptive strategy selection, modeled as a multi-armed bandit problem~\cite{lattimore2020bandit,agrawal2011ts,cavenaghi2021nonstationary} to improve efficiency and reduce redundant requests. (2) We design a context-aware vulnerability detector that leverages LLM reasoning to interpret responses and identify diverse vulnerability categories beyond static rule-based detection. (3) We conduct extensive experimental evaluation on open-source and benchmark GraphQL APIs, demonstrating that \pq~achieves higher coverage and detects more context-dependent vulnerabilities than existing tools.

\newenvironment{editblock}[1]{\begingroup\color{#1}}{\endgroup}
\section{BACKGROUND \& RELATED WORK}

Modern web applications increasingly adopt GraphQL for its flexible and efficient data fetching model. While this paradigm simplifies client–server interaction, it also introduces a new class of security challenges that differ from those found in traditional REST APIs. In this section, we first review the fundamentals of GraphQL, then summarize its known vulnerability classes, and finally discuss prior work on API testing, GraphQL security analysis, large language model (LLM)–assisted fuzzing, prompt engineering, and adaptive test strategy selection.

\subsection{GraphQL Fundamentals}

GraphQL provides a structured and flexible alternative to traditional REST APIs, offering three key features \cite{graphql_security_page_2025,aws_graphql_rest_2024,andersson2021rest_graphql_grpc}: 

\begin{itemize}[leftmargin=*, label=$\diamond$]

\item \textit{Data as a graph:} GraphQL organizes data as a graph of interconnected objects, allowing clients to retrieve all required information in a single request and mitigating both under-fetching and over-fetching.  
\item \textit{Strong typing:} Each GraphQL API exposes a schema that defines data types (objects) and operations (queries and mutations). This strong type system ensures predictable responses, simplifies error handling, and requires clients to explicitly specify which fields to return.  
\item \textit{Single endpoint.} All requests are processed through a unified endpoint, providing a consistent interface for both data retrieval and modification.

\end{itemize}
\begin{figure*}[t]
  \centering
    \includegraphics[width=0.8\linewidth]{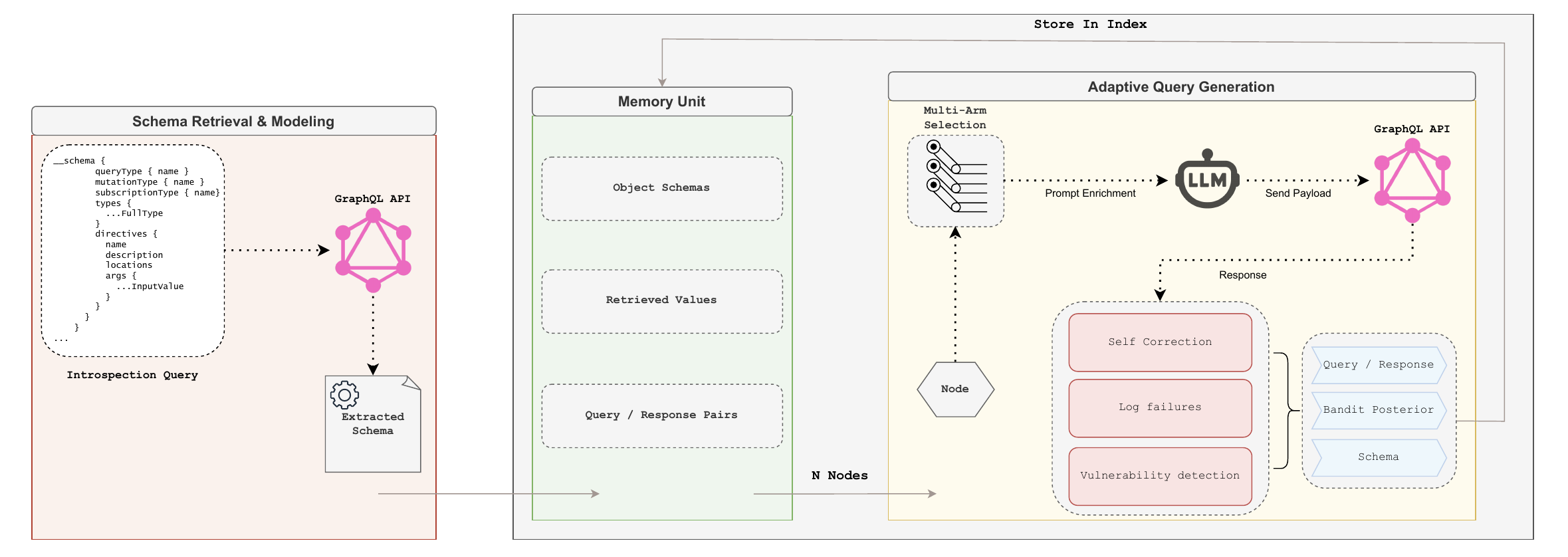}
    \caption{ workflow of \pq.}
  \label{fig:pipeline}
\end{figure*}

A GraphQL schema defines two primary categories of data types: \textit{scalars} and \textit{objects}.  
Scalars represent atomic values such as \texttt{Int}, \texttt{Float}, \texttt{String}, \texttt{Boolean}, and \texttt{ID}.  
Objects, on the other hand, are user-defined entities composed of multiple fields, which may themselves be scalars, objects, or lists of other types.
This nested composition creates a rich graph structure that captures relationships among entities and allows clients to traverse and query linked data seamlessly.

Clients interact with GraphQL APIs through two main operation types:
\begin{itemize}[leftmargin=*, label=$\diamond$]
    \item \textit{Queries} retrieve data from the server, functioning similarly to HTTP \texttt{GET} requests in REST APIs. A special form, the \textit{introspection query}, allows clients to inspect the schema and obtain metadata about available types and operations.
    \item \textit{Mutations} modify data on the server, corresponding to creation, update, or deletion actions.
\end{itemize}

Both queries and mutations enable fine-grained field selection, ensuring clients request exactly the data they need. Each field may also take arguments, allowing deeply nested and parameterized requests. A GraphQL operation is first parsed and validated against the schema, then executed by resolving each field independently. This layered execution model, together with features like schema introspection, makes GraphQL powerful but also complex to test securely.

\subsection{GraphQL Vulnerabilities}
\label{sec:graphql_vulnerabilities}

GraphQL provides powerful ways to query and manage data, but its flexibility also brings new types of security risks. Some of these problems are similar to common web application issues, such as broken access control, injection flaws, and misconfigurations, while others are specific to how GraphQL is designed. These GraphQL-specific issues expand the attack surface and need special attention. We group them into three main categories:

\begin{itemize}[leftmargin=*, label=$\diamond$]

\item \textit{Query Abuse Vulnerabilities.}  Because GraphQL allows users to send highly flexible queries, attackers can take advantage of this feature to overload or explore the system. A common example is misuse of the \textit{introspection query}, which reveals the entire schema, including all types, fields, and relationships. This information helps attackers craft more targeted and harmful queries. GraphQL is also exposed to \textit{Denial of Service (DoS)} attacks caused by deeply nested or repetitive queries that consume large amounts of server resources, leading to slowdowns or outages.

\item \textit{Injection Vulnerabilities.}  GraphQL inputs can be an entry point for injection attacks if they are not properly checked or sanitized. The most serious case is \textit{SQL Injection}, where unsafe user inputs are added to database queries, allowing data theft or manipulation. Other examples include \textit{Path Injection} and \textit{Cross-Site Scripting (XSS)}. These can result in stolen sessions, leaked data, or malicious code running in the browser. Weak input handling in GraphQL therefore puts both the backend and users at risk.

\item \textit{Access Control Vulnerabilities.}  Access control problems occur when a GraphQL API fails to properly restrict what data users can access. A typical example is \textit{Insecure Direct Object Reference (IDOR)}, where attackers change object identifiers to view or modify restricted data. Another example is \textit{batched attacks}, where multiple operations are combined in one request to bypass individual security checks. Fixing such issues is difficult because GraphQL schemas often include many linked fields and relationships. Testing for these problems requires context-aware and dependency-based approaches that can understand how queries, mutations, and objects interact, helping to uncover access control flaws that traditional testing might miss.

\end{itemize}

\subsection{Related Work}

\noindent {\bf GraphQL Security Testing.} 
Security testing for GraphQL has evolved from basic black-box approaches to more sophisticated schema-aware methods. 
Early community tools such as {GraphQL-Cop}~\cite{graphqlcop}, {GraphCrawler}~\cite{graphcrawler}, {CrackQL}~\cite{crackql}, and Schemathesis~\cite{hatfield_dodds_2021_schemathesis} rely on introspection to generate single-request mutations but lack systematic reasoning about dependencies between queries and mutations.  
{EvoMaster} extends evolutionary and random testing to GraphQL APIs, supporting both black-box and white-box modes to broaden coverage~\cite{belhadi2023evomaster,arcuri2021evomaster}. 
Industry-standard scanners such as {OWASP ZAP}~\cite{zap2024} and {BurpSuite}~\cite{burpsuite} include GraphQL modules (e.g., Burp’s Auto GQL Scanner) but primarily perform generic payload fuzzing and introspection-based attacks rather than exploring multi-step workflows.  
{GraphQLer}~\cite{graphqler2025} introduced the first context-aware GraphQL security testing framework. It analyzes the schema to infer producer–consumer dependencies among queries and mutations, constructs a dependency graph, and generates realistic chained payloads. This approach improves schema coverage and discovers previously unknown vulnerabilities compared with earlier fuzzers, motivating the need for dependency reasoning in GraphQL security testing.

\vspace{1mm}

\noindent {\bf LLM-Assisted Fuzzing.}
In dynamic analysis, fuzzing remains one of the most effective ways to automatically find software bugs. 
Coverage-guided fuzzers such as AFL++~\cite{aflplusplus2020} and libFuzzer~\cite{libfuzzer}, along with large-scale efforts like OSS-Fuzz~\cite{ossfuzz}, have discovered thousands of vulnerabilities. 
However, these approaches often struggle with highly structured or constrained inputs and with reaching deep execution paths. 
To address these challenges, recent work combines LLMs with fuzzing and related dynamic techniques. 
{Fuzz4All}~\cite{xia2023fuzz4all} uses LLMs for input generation and mutation across multiple formats, maintaining diversity through an autoprompting loop and achieving higher coverage than traditional fuzzers.  {ELFuzz}~\cite{chen2025elfuzz} enhances generation-based fuzzers through LLM-driven synthesis over input spaces, achieving higher coverage and revealing real-world bugs. 
Recent studies have explored machine learning–based detection of malicious GraphQL queries. For instance, one line of work applies deep reinforcement learning to identify denial-of-service patterns in GraphQL APIs \cite{ieee10579536}, while another investigates the use of large language models, sentence transformers, and convolutional neural networks for malicious query detection \cite{perera2025enhancing}. There has also been work to use LLMs in text-to-GraphQL queries \cite{graphqlquerygenerationkesarwani}; however, the model does not target specific API schemas. To the best of our knowledge, no prior work has applied LLM-assisted fuzzing to GraphQL, whose schema-rich, multi-operation design introduces unique challenges. 
\pq~is the first to address this gap by leveraging LLM reasoning to guide test generation and vulnerability detection in GraphQL APIs.

\section{METHODOLOGY}

\pq~is a modular LLM-driven GraphQL fuzzing framework organized around a closed-loop pipeline (Figure~1).
At its core, \pq~integrates an LLM-based query generator with schema introspection, semantic retrieval, adaptive arm selection, and self-correction modules. This architecture constrains the model’s generation space, maintaining validity and fuzzing relevance while adapting to feedback dynamically. The key contribution lies in coupling schema-aware reasoning with adaptive prompting to produce valid, diverse, and feedback-driven queries.

Rather than relying on static mutation operators, \pq~refines prompts with schema fragments, retrieved traces, and error signals, making each generation guided by prior results and schema insights. The system runs as a closed-loop cycle: schema modeling constrains the search space, the bandit-based selector chooses a prompting strategy, retrieval grounds prompts in real execution history, and self-correction incorporates prior errors. Prompt construction then assembles these elements into an evidence-gated LLM input; executed queries update the schema, memory, and bandit posteriors, completing the loop and allowing iterative refinement of validity, diversity, and coverage.

\subsection{Schema Modeling}
\noindent
The \pq~pipeline begins with a standard GraphQL introspection query against the target API. Introspection reveals the complete schema, which includes queries, mutations, input parameters, and return types. \pq~parses this output into a structured intermediate representation that organizes operations, argument specifications, and object definitions. To facilitate reuse and downstream automation, the schema is serialized into lightweight YAML files, separating queries, mutations, and type definitions. This schema map provides a blueprint of the API before any fuzzing or query generation takes place. 

Unlike prior tools that treat introspection results as flat listings, \pq~recursively follows links between objects to capture nested and cross-referenced types. This yields a graph-structured view of responses, enabling the system to reason about both top-level operations and the shape of deeply nested return values. Maintaining this normalized schema representation allows \pq~to generate queries that are syntactically valid, semantically consistent, and structurally diverse, while avoiding common issues such as type mismatches or missing arguments.

\subsection{Adaptive Query Generation}

\noindent \textbf{Adaptive Arm Selection.} Query generation is framed as a multi-armed bandit problem \cite{lattimore2020bandit}, where each arm represents a distinct prompting strategy defined by four key parameters. The \textit{Schema} parameter determines whether the full GraphQL schema is included in the LLM prompt to guide query construction. The \textit{Arg Mode} parameter controls how arguments are generated: \textit{known} reuses previously successful parameter values from the RAG context, \textit{real} synthesizes realistic type-appropriate literals, and \textit{nulls} tests optional fields with null values. The \textit{Depth} parameter limits the nesting level of GraphQL selections to balance complexity and validity, while the \textit{top-k} parameter specifies how many similar examples from the RAG system are retrieved as contextual grounding (see Section~\ref{sec:setup} for configuration details).

Since the effectiveness of each strategy depends on the endpoint’s evolving behavior, the bandit formulation enables \pq~to dynamically allocate preference toward high-performing arms while still exploring alternatives. To balance exploration and exploitation, \pq~employs Thompson Sampling \cite{agrawal2011ts}, rewarding arms only when generated queries both {\it (i)} return HTTP 200 responses and {\it (ii)} expand coverage. To account for non-stationary environments, rewards are exponentially discounted so that outdated strategies are gradually down-weighted. This adaptive mechanism prevents over-commitment to a single strategy while progressively amplifying those that yield meaningful coverage.

\vspace{1mm}

\noindent {\textbf{Retrieval-Augmented Generation.} To mitigate the problem of hallucinated or invalid query values, \pq~integrates retrieval-augmented generation (RAG) \cite{lewis2020rag}. All prior queries and responses are embedded into a FAISS index \cite{douze2024faiss}. During generation, \pq~retrieves the top-$k$ (where $k$ is equal to 0, 3, or 5, depending on the selected ARM) most semantically relevant traces and injects them into the prompt.}

{This retrieval memory grounds the LLM in real execution history, improving syntactic fidelity, reducing repetition, and promoting diversity by surfacing alternative query structures.}

\vspace{1mm}

\noindent {\textbf{Prompt Engineering.} Prompt design is key to guiding query generation. \pq~adopts four design goals: prompts must be evidence-gated, deterministic, schema-constrained, and context-aware. At runtime, prompt $P$ is automatically assembled from five modular components: $P = [B \,\|\, S \,\|\, R \,\|\, E \,\|\, D]$:}

\begin{itemize}[leftmargin=*, label=$\triangleright$]    
\item $B$: Basic restricting header (evidence-gating).
    \item $S$: Schema fragments from introspection
    \item $R$: Retrieved execution examples (RAG)
    \item $E$: Prior error--query pairs (self-correction)
    \item $D$: Strategy-specific directives (bandit arm)
\end{itemize}

{This modular structure, illustrated in Figure 2, ensures that exploration happens primarily through $D$ (arm choice), while $S$, $R$, and $E$ stabilize performance across strategies.}

\begin{figure}[t]
  \centering
  \includegraphics[width=\linewidth]{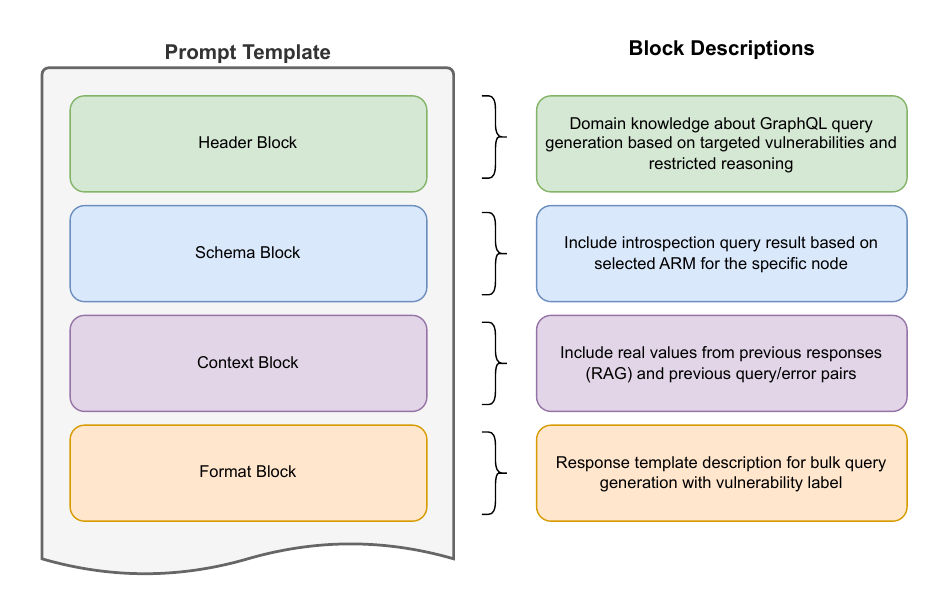}
\caption{structure of curated prompt and enrichment sections.}
  \label{fig:prompt_structure}
\end{figure}

\subsection{Execution and Feedback}

\textbf{Self Correction.} Failed queries are not discarded. Instead, \pq~explicitly records schema errors and associates them with the queries that caused them. In subsequent generations, these error–query pairs are injected into the prompt as corrective signals.

This error-guided refinement turns schema violations into supervision, steering the model away from repeated mistakes and accelerating convergence toward valid, schema-compliant queries.

\vspace{1mm}

\noindent \textbf{Context-Aware Vulnerability Detection.} To assess security impact, \pq~uses an LLM to analyze responses in context, rather than relying on predefined signatures or static rules. Each response, together with its execution metadata such as status codes and error messages, is transformed into a structured analysis prompt that directs the LLM to identify evidence of vulnerabilities, including injection flaws, access-control bypasses, and information disclosure. The resulting analyses are parsed into JSON records encoding vulnerability type, severity, confidence, and evidence. These records are stored and aggregated, yielding both granular findings and system-level trends. By conditioning on schema knowledge and execution context, \pq~generalizes across diverse GraphQL APIs and uncovers subtle, context-dependent flaws that rule-based detectors overlook.

\subsection{Closed-Loop Integration}
{Together, these phases form a closed-loop fuzzing cycle in which schema knowledge constrains the LLM, adaptive generation explores diverse query strategies, and execution feedback continuously refines both prompts and strategy selection. This iterative design enables \pq~to expand coverage systematically while maintaining robustness and precision in vulnerability discovery.}

\section{EVALUATION}

We implement the design and experiments in an open-source repository\footnote{https://github.com/SLL288/prediql}. Our experiment is structured to investigate the following research questions: \textbf{RQ1.} How can LLMs be guided to synthesize valid yet adversarial GraphQL queries that systematically expand schema coverage compared to schema-only or random fuzzing approaches? \textbf{RQ2.} How does prompt engineering and context enrichment contribute to improving schema coverage? \textbf{RQ3.} To what extent can this pipeline discover meaningful GraphQL vulnerabilities over existing rule-based methods?

\subsection{Experimental Setup}
\label{sec:setup}
\noindent \textbf{API Selection.} In our experimental evaluation, we employed a diverse spectrum of APIs, including open-source and openly hosted options. For open-source APIs, we obtained the backend code, established a self-hosted GraphQL server, and conducted tests. For openly hosted APIs, we utilized free-to-use reference APIs accessible on GitHub and conducted direct testing. The selected APIs are shown in Table \ref{tab:apis}.

\begin{table}[t]
\small
\centering
\caption{GraphQL APIs used in baseline testing.}

\begin{tabular}{lccc}
\toprule
\textbf{API} & \textbf{\#Queries} & \textbf{\#Mutations} & \textbf{\#Objects} \\
\midrule
{\np UserWallet} \cite{userwallet_api}     & 11  & 15 & 5    \\
{\np Countries} \cite{countries_api}       & 6   & 0  & 5    \\
{\np Rick\&Morty} \cite{rickmorty_api}     & 9   & 0  & 7    \\
{\np GraphQLZero} \cite{graphqlzero_api}     & 13  & 19 & 18   \\
{\np EHRI} \cite{ehri_api}            & 19  & 0  & 46   \\
{\np TCGDex} \cite{tcgdex_api}        & 6   & 0  & 12   \\
\bottomrule
\end{tabular}

\label{tab:apis}
\end{table}

\begin{table}[b]
\centering
\caption{Technical specifications of LLMs used for GraphQL testing.}
\resizebox{\columnwidth}{!}{%
\begin{tabular}{lcccc}
\toprule
\textbf{Model} & \textbf{Developer} & \textbf{Year} & \textbf{Size} & \textbf{Focus} \\
\midrule
{\zw LLaMA 3.1}     & Meta           & 2024 & 8B & Open-source, efficient inference \\
{\zw Gemini 2.5 Flash} & Google DeepMind& 2025 & Undisclosed & cost-efficient reasoning \\
{\zw GPT-5 Mini}      & OpenAI         & 2025 & Undisclosed & optimized for speed \\
{\zw DeepSeek R1}      & DeepSeek AI    & 2025 & 671B & Reasoning and efficiency \\
\bottomrule
\end{tabular}%
}
\label{tab:model-comparison}
\end{table}
\vspace{1mm}
\noindent {\bf Baselines.} We evaluate \pq~against four representative baselines that include both general-purpose and GraphQL-specific testing frameworks.

\begin{itemize}[leftmargin=*, label=$\diamond$]
\item \textsf{EvoMaster}. The only prior academic framework supporting GraphQL testing in both white-box and black-box modes. We use its \emph{black-box} configuration for fair comparison. \textsf{EvoMaster} generates and sends GraphQL queries and mutations automatically, but employs evolutionary heuristics to mutate payloads dynamically and explore a wider response space~\cite{belhadi2023evomaster,arcuri2021evomaster,belhadi2023graphql}.

\item \textsf{ZAP}. An open source black-box vulnerability scanner maintained by the Open Web Application Security Project (OWASP)~\cite{zap2024}. Although originally designed for traditional web applications, it includes a module for GraphQL testing that performs introspection-based payload generation and common attack simulations.

\item \textsf{BurpSuite}. A widely used commercial web security platform developed by PortSwigger~\cite{burpsuite}. For GraphQL testing, we enable the \textit{Auto GQL Scanner} extension~\cite{burpsuite_scanner}, which automatically identifies GraphQL endpoints and injects predefined payloads to detect vulnerabilities such as injection and schema disclosure.

\item \textsf{GraphQLer}. A recent context-aware GraphQL security testing framework~\cite{graphqler2025}. It constructs a dependency graph that captures producer–consumer relationships between queries and mutations, allowing the generation of realistic chained requests. GraphQLer demonstrates the benefit of dependency reasoning for GraphQL security testing and serves as the strongest baseline in our comparison.

\end{itemize}

\begin{table}[b]
\centering
\caption{Parameters defining adaptive arms in \textsc{PrediQL}’s prompt generation.}
\label{tab:bandit_config}
\small
\begin{tabular}{lcccc}
\toprule
\textbf{Arm Name} & \textbf{Schema} & \textbf{Arg Mode} & \textbf{Depth} & \textbf{Top-k} \\
\midrule
\texttt{schema\_min\_known} & True  & known & 1 & 3 \\
\texttt{schema\_min\_real}  & True  & real  & 1 & 3 \\
\texttt{schema\_mod\_known} & True  & known & 2 & 5 \\
\texttt{noschema\_min\_known} & False & known & 1 & 3 \\
\texttt{noschema\_min\_real}  & False & real  & 1 & 0 \\
\texttt{schema\_min\_nulls} & True  & nulls & 1 & 3 \\
\texttt{schema\_deep\_known} & True  & known & 3 & 5 \\
\texttt{schema\_deep\_real}  & True  & real  & 3 & 5 \\
\bottomrule
\end{tabular}
\end{table}

\noindent \textbf{Model Selection.}
For evaluation, we selected four LLMs spanning different developers, sizes, and design philosophies, as summarized in Table~\ref{tab:model-comparison}. Specifically, we included \textsf{{\zw LLaMA 3.1}} from Meta~\cite{meta_llama_3_1_8b} as a representative open source model, \textsf{{\zw Gemini 2.5}} from Google DeepMind~\cite{deepmind_gemini_2_5_flash} as a lightweight and optimized variant of the Gemini family, \textsf{GPT-5 Mini} from OpenAI~\cite{gpt5_mini_comparison} as a smaller yet efficient version of the GPT-5 series, and \textsf{{\zw DeepSeek R1}} from DeepSeek AI~\cite{deepseek_r1_comparison}, an open source model that emphasizes reasoning and efficiency. This diverse selection allows us to cover both open source and proprietary approaches, models optimized for speed as well as those focused on reasoning capabilities.

\begin{table*}[!t]
\centering
\begin{minipage}{0.6\textwidth}
\caption{Comparison of \pq~and baseline methods in terms of coverage performance. Tests that could not be executed are marked as \texttt{FAILED}.}

\begin{tabular}{lccccc}
\toprule
\textbf{API} & \textbf{ZAP} & \textbf{Burp} & \textbf{EvoMaster} & \textbf{GraphQLer} & \textbf{\pq} \\
\midrule
{\np UserWallet} & 50.00\% & 7.69\% & 61.54\% & 92.31\% & \textbf{96.15\% (\textcolor{OliveGreen}{+3.84\%})} \\ {\np Countries} & 33.33\% & 50.00\% & 50.00\% & 50.00\% & \textbf{100\% (\textcolor{OliveGreen}{+50.00\%})} \\ {\np Rick\&Morty} & 33.33\% & 0.00\% & 66.67\% & 66.67\% & \textbf{100\% (\textcolor{OliveGreen}{+33.33\%})} \\ {\np GraphQLZero} & 93.75\% & 93.75\% & 71.88\% & 93.75\% & \textbf{100\% (\textcolor{OliveGreen}{+6.25\%})} \\{\np EHRI} & 10.53\% & 0.00\% & 84.21\% & 94.74\% & \textbf{100\% (\textcolor{OliveGreen}{+5.26\%})} \\ 
{\np TCGDex} & 66.67\% & 33.33\% & 100\% & 100\% & \textbf{100\% (\textcolor{Gray}{+0.00\%})} \\
\bottomrule
\end{tabular}%

\label{tab:prediql-comparison}
\end{minipage}
\end{table*}

\vspace{1mm}
\noindent \textbf{Bandit Configuration Details.} For experimental evaluation, we instantiated eight distinct bandit arms, each representing a specific prompting strategy defined by four parameters: \texttt{Schema}, \texttt{Arg Mode}, \texttt{Depth}, and \texttt{Top-k}. Table \ref{tab:bandit_config} summarizes these configurations.

Each configuration corresponds to a balance between reliability and exploration. Conservative arms (\textit{e.g.}, \texttt{schema\_min\_known}) prioritize syntactic validity by reusing known parameter values and shallow nesting. Aggressive arms (\textit{e.g.}, \texttt{schema\_deep\_real}) promote deeper traversal and the synthesis of new argument values. The non-schema variants probe the LLM’s ability to generalize without explicit schema guidance. The bandit dynamically reallocates probability mass toward arms that maximize successful, coverage-expanding executions.

\vspace{1mm}
\noindent \textbf{Evaluation Metric.}
To address RQ1 and RQ2, we define the following coverage metric to evaluate GraphQL API testing:

$$ \mathbf{\textit{Coverage}} = \frac{\#\text{Unique Successful Responses}}{\#\text{Unique Nodes}} $$

\noindent where \textit{coverage} is the fraction of schema nodes that return valid, error-free data rather than just an HTTP 200 status. This metric reflects how much of the GraphQL schema is actually exercised by successful queries, providing a more accurate view of the API’s reliability and robustness.

\subsection{Schema Coverage (RQ1)}
  Table \ref{tab:prediql-comparison} compares the API coverage achieved by \pq~against four baseline tools: \textsf{ZAP}, \textsf{BurpSuite}, \textsf{EvoMaster}, and \textsf{GraphQLer}. Across all evaluated APIs, \pq~consistently attains the highest or near-highest coverage, demonstrating its ability to exercise a larger portion of the GraphQL schema through valid, data-returning queries. Traditional black-box scanners such as \textsf{ZAP} and \textsf{BurpSuite} achieve limited coverage because they lack awareness of GraphQL’s hierarchical structure and query dependencies. \textsf{EvoMaster} performs better by generating dynamic requests, but still falls short on complex schemas. GraphQLer improves coverage by incorporating schema context, yet \pq~surpasses it in nearly all cases by integrating retrieval-augmented reasoning and adaptive query generation. Note that the reported \pq~results correspond to its configuration with the best-performing LLM.

  Table \ref{tab:apis-models} presents the performance of tested large language models across different GraphQL APIs. Overall, all models achieve high coverage on simpler or well-structured APIs such as {\np Countries}, {\np Rick \& Morty}, {\np GraphQLZero}, and {\np TCGDex}, indicating that these schemas are easier to navigate and query successfully. However, differences emerge on more complex or noisy schemas such as {\np UserWallet} and {\np EHRI}. {\zw GPT-5 Mini} and {\zw Gemini 2.5} consistently produce the most stable and complete results, suggesting stronger schema understanding and query adaptation capabilities. {\zw LLaMA 3.1} performs comparably but occasionally misses certain paths. In contrast, {\zw DeepSeek R1} positions itself as the second-most proficient model on specific APIs, closely following {\zw GPT-5 Mini}. These findings suggest that stronger reasoning models maintain high coverage under schema complexity, indicating better adaptability and understanding of GraphQL structures.   

\begin{rqbox}[Conclusion for RQ1]
  
  \pq~consistently outperforms all baseline tools in schema coverage across diverse APIs. Its context-aware input inference and semantic reasoning enable more accurate and comprehensive query generation. On average, \pq~achieves a 16\% improvement with a maximum improvement of 50\% in schema coverage over the second-best model.  


\end{rqbox}

\subsection{Prompt Engineering Impact (RQ2)}

Prompt engineering is central to steering the reasoning capabilities of LLMs in improving schema coverage and discovering vulnerabilities in APIs. To address RQ2, we conduct an ablation study that isolates the contribution of each prompt enrichment component. This analysis quantifies how each module influences schema coverage and overall testing effectiveness.  The corresponding configurations are detailed below.

\begin{itemize}[leftmargin=*, label=$\diamond$]
\item \textbf{\textsc{PrediQL-BASE}}. The {baseline} configuration provides only the minimal schema context and expected response format. It guides the model to produce syntactically valid GraphQL queries.


\item \textbf{\textsc{PrediQL-AQG}}. Building on the baseline, the adaptive query generation configuration integrates both multi-armed bandit selection and retrieval-augmented generation, which is essential for \textit{known values} ARM setting to enable adaptive and context-aware query synthesis.


\item \textbf{\textsc{PrediQL-SCL}}. The self-correction feedback loop configuration enhances the baseline prompt, adding an error-aware refinement cycle. Failed or invalid queries are logged with their corresponding error messages and reinjected into subsequent prompts as a corrective context.

\item \textbf{\textsc{PrediQL}}. The full pipeline, combining adaptive query generation, retrieval augmentation, and self-correction into a single closed-loop system.
\end{itemize}

Table~\ref{tab:ablation} evaluates the contribution of individual prompt engineering components: the base prompt (\textsc{BASE}), the self-correction loop (\textsc{PrediQL-SCL}), and adaptive query generation (\textsc{PrediQL-AQG}). Contributions are computed for models that experienced enhancements from PrediQL-Base to PrediQL.
Across models and APIs, the incremental addition of these modules consistently improves schema coverage, confirming that each contributes complementary benefits.  
\textsc{PrediQL-BASE} alone often yields limited coverage, particularly in complex schemas such as {\np UserWallet} and {\np EHRI}, where naive prompting fails to satisfy nested or dependent field constraints.  
Introducing the self-correction loop (\textsc{PrediQL-SCL}) markedly reduces repeated schema violations, increasing coverage by 10–25\% depending on the model capacity.  
Adaptive query generation (\textsc{PrediQL-AQG}) provides an additional boost to APIs that require realistic parameter inference, increasing coverage by as much as 26\% on {\np EHRI} and 15\% on {\np GraphQLZero} for \texttt{\zw Gemini 2.5}.  
When both mechanisms are combined in \pq, coverage approaches or reaches 100\% in almost all APIs and models.  

Smaller open-source models such as {\zw LLaMA 3.1} follow the same pattern but exhibit slightly higher variance, reflecting reduced stability in long-context reasoning.  
Larger models ({\zw GPT-5 Mini}, {\zw DeepSeek R1}) demonstrate less sensitivity to prompt enrichment, suggesting that reasoning-oriented architectures benefit more from the feedback of retrieval and correction than from raw scale alone.  In general, ablation confirms that the design gains of \pq stem from the synergy of retrieval grounding, adaptive prompting, and iterative self-correction, rather than from model size alone.

\begin{table*}[t]
\centering
\begin{minipage}{0.6\textwidth}
\caption{Coverage achieved by different language models across GraphQL APIs.}
\footnotesize
\resizebox{\columnwidth}{!}{%
\begin{tabular}{lcccc}
\toprule
\textbf{API} & \textbf{{\zw LLaMA 3.1}} & \textbf{{\zw Gemini 2.5}} & \textbf{GPT-5 Mini} & \textbf{{\zw DeepSeek R1}} \\
\midrule
{\np UserWallet}    & 88.46\% & 96.15\% & 96.15\% & 88.46\% \\
{\np Countries}     & 100\% & 100\% & 100\% & 100\% \\
{\np Rick\&Morty}   & 100\% & 100\% & 100\% & 100\% \\
{\np GraphQLZero}   & 100\% & 100\% & 100\% & 100\% \\
{\np EHRI}          & 78.94\% & 100\% & 100\% & 100\% \\
{\np TCGDex}        & 100\% & 83.33\% & 100\% & 100\% \\
\bottomrule
\end{tabular}%
}
\label{tab:apis-models}
\end{minipage}
\end{table*}

\begin{table*}[b]
\centering
\caption{Ablation study on prompt engineering components. \textsc{BASE}, \textsc{AGQ}, and \textsc{SCL} denote \textsc{PrediQL-BASE}, \textsc{PrediQL-AGQ}, and \textsc{PrediQL-SCL}, respectively.}
\resizebox{\textwidth}{!}{
\begin{tabular}{lcccccccccccccccc}
\toprule
\textbf{API} & 
\multicolumn{4}{c}{\textbf{{\zw GPT-5 Mini}}} &
\multicolumn{4}{c}{\textbf{{\zw Gemini 2.5}}} &
\multicolumn{4}{c}{\textbf{{\zw DeepSeek R1}}} &
\multicolumn{4}{c}{\textbf{{\zw LLaMA 3.1}}} \\
\cmidrule(lr){2-5} \cmidrule(lr){6-9} \cmidrule(lr){10-13} \cmidrule(lr){14-17}
 & \textbf{BASE} & \textbf{SCL} & \textbf{AQG} & \textbf{\pq}
 & \textbf{BASE} & \textbf{SCL} & \textbf{AQG} & \textbf{\pq}
 &  \textbf{BASE} & \textbf{SCL} & \textbf{AQG} & \textbf{\pq}
 & \textbf{BASE} & \textbf{SCL} & \textbf{AQG} & \textbf{\pq} \\
\midrule
{\np UserWallet}   & 19.23\% & 38.46\% & 61.53\% & 96.15\%  & 19.23\% & 26.92\% & 65.38\% & 96.15\%& 19.23\% & 30.76\% & 65.38\% & 96.15\%& 38.46\% & 42.30\% & 84.61\% & 88.46\%\\
{\np Countries}    & 100\% & 100\% & 100\% & 100\% & 100\% & 100\% & 100\% & 100\% & 100\% & 100\% & 100\% & 100\% & 100\% & 100\% & 100\% & 100\% \\
{\np Rick\&Morty}  & 100\% & 100\% & 100\% & 100\% & 100\% & 100\% & 100\% & 100\% & 100\% & 100\% & 100\% & 100\% & 100\% & 100\% & 100\% & 100\% \\
{\np GraphQLZero}  & 81\% & 100\%  & 87.5\% & 100\% 
             & 81\% & 100\%  & 91\%  & 100\% 
             & 100\% & 100\% & 100\% & 100\% 
             & 90\% & 100\% & 96.87\% & 100\% \\
{\np EHRI}         & 100\% & 100\% & 100\% & 100\% & 74\% & 74\% & 100\% & 100\% & 52.63\% & 52.63\% & 84.21\% & 100\% & 52\% & 100\% & 100\%  & 78.94\%  \\
{\np TCGDex}       & 100\% & 100\% & 100\% & 100\% & 83\% & 83\% & 83\% & 83\% & 100\% & 100\% & 100\% & 100\% & 100\% & 100\% & 100\% & 100\% \\
\bottomrule
\end{tabular}
}
\label{tab:ablation}
\end{table*}

\begin{rqbox}[Conclusion for RQ2]
Overall, prompt engineering significantly enhances \pq's ability to achieve broader and more accurate schema coverage.
By integrating adaptive query generation and self-correction, the system effectively adapts, learns, and refines its queries over iterations.
This synergy results in a more intelligent, reliable, and efficient API exploration process.
\end{rqbox}

\begin{table*}[t]
\centering
\caption{Comparison of vulnerability detection performance between GraphQLer and \textsc{PrediQL} variants.}

\resizebox{\textwidth}{!}{
\begin{tabular}{l|cc|cc|cc|cc|cc}
\toprule
\multirow{1}{*}{\textbf{API}} & \multicolumn{2}{c|}{\textbf{GraphQler}} & \multicolumn{2}{c|}{\textbf{\pq~({\zw LLaMA 3.1})}} & \multicolumn{2}{c|}{\textbf{\pq~({\zw Gemini 2.5})}} & \multicolumn{2}{c|}{\textbf{\pq~({\zw GPT-5 Mini)}}} & \multicolumn{2}{c}{\textbf{\pq~({\zw DeepSeek R1})}} \\
\cmidrule(lr){2-3} \cmidrule(lr){4-5} \cmidrule(lr){6-7} \cmidrule(lr){8-9} \cmidrule(lr){10-11}
 & Vulnerabilities & Categories & Vulnerabilities & Categories & Vulnerabilities &  Categories & Vulnerabilities & Categories & Vulnerabilities & Categories \\
\midrule
{\np UserWallet}    & 26 & 7 & 31 & 11 & 41 & 7 & 20 & 6 & 34 & 8 \\
{\np Countries}     & 6 & 2 & 7 & 3 & 9 & 2 & 9 & 4 & 7 & 3 \\
{\np Rick\&Morty}   & 12 & 3 & 10 & 10 & 13 & 4 & 11 & 4 & 14 & 6 \\
{\np GraphQLZero}   & 37 & 8 & 35  & 7 & 37 & 7 & 44 & 6 & 34 & 7 \\
{\np EHRI}          & 11 & 3 & 15 & 12 & 21 & 2 & 26 & 2 & 3 & 3 \\
{\np TCGDex}        & 6 & 1 & 7 & 1 & 10 & 2 & 8 & 2 & 7 & 2 \\
\bottomrule

\end{tabular}%
}

\label{tab:api-detection}
\end{table*}

\subsection{Vulnerability Detection (RQ3)}

Table~\ref{tab:api-detection} summarizes the vulnerability detection results across \textsf{GraphQLer} and the different configurations of \textsc{PrediQL}. Among existing GraphQL testing tools, only \textsf{GraphQLer} includes a built-in vulnerability detection module, while others, such as \textsc{EvoMaster} and \textsc{ZAP}, primarily focus on coverage measurement or payload fuzzing. Therefore, \textsf{GraphQLer} serves as the most relevant baseline for assessing the detection capacity.

Across all evaluated APIs, \textsc{PrediQL} consistently identifies a greater number and a wider range of vulnerabilities. While \textsf{GraphQLer} mainly exposes schema-level and input validation flaws, \textsc{PrediQL} leverages retrieval-augmented reasoning and adaptive arm selection to detect deeper logic- and context-dependent weaknesses such as HTML injection, SSRF, and OS command injection. The variants \textsc{PrediQL} -Gemini and \textsc{PrediQL} -GPT-5 achieve the highest detection counts, improving unique findings by 20-40\% on complex benchmarks such as {\np UserWallet} and {\np GraphQLZero}. In general, these results confirm that LLM-guided reasoning substantially enhances vulnerability discovery beyond static or heuristic testing baselines.

Qualitative analysis shows that \textsc{PrediQL} detects a broader spectrum of vulnerability categories compared to \textsf{GraphQLer}, including deeper logic and context-dependent flaws. Its reasoning traces link each issue to its execution context (e.g., leaked variables or inconsistent authorization responses), enabling precise, evidence-based triaging.

\begin{rqbox}[Conclusion for RQ3]
\pq~substantially advances vulnerability discovery beyond rule-based baselines.
Its context-aware reasoning enables the LLM to correlate schema structure, response semantics, and execution traces, revealing logic and injection flaws that static or signature-driven tools overlook.
Across all evaluated APIs, \pq~achieves broader and deeper vulnerability coverage, demonstrating that adaptive, retrieval-guided analysis is essential to uncovering complex security weaknesses in GraphQL APIs.
\end{rqbox}

\section{Discussion}

Our evaluation demonstrates that \pq~consistently outperforms all existing GraphQL testing frameworks. This section discusses key observations, broader implications, and open research directions.

\vspace{1mm}
\noindent \textbf{Impact of Model Size.} Larger models such as {\zw GPT-5 Mini} and {{\zw DeepSeek R1}} achieved higher semantic coherence and reasoning stability, while smaller open-source models (e.g., {{\zw LLaMA 3.1}}) remained competitive at a fraction of the computational cost. This highlights a practical trade-off between reasoning depth and efficiency. When equipped with retrieval memory and adaptive prompting, lightweight models can approximate the performance of proprietary ones. A promising direction is to adopt hybrid configurations, using larger models for seed generation and schema understanding, followed by smaller models for iterative fuzzing, to balance throughput, coverage, and cost.

\vspace{1mm}

\noindent \textbf{Implications Beyond GraphQL.} The introduced mechanisms in \pq, adaptive arm selection, self-corrective prompting, and retrieval-grounded reasoning, are not specific to GraphQL. These ideas extend naturally to other structured interface testing domains such as REST, gRPC, and JSON-RPC \cite{jsonrpc_specification}. More broadly, \pq~demonstrates that retrieval-augmented reasoning and bandit-driven exploration can complement traditional coverage-guided and evolutionary fuzzing. Integrating symbolic reasoning or static program analysis into such adaptive loops may bridge the gap between semantic understanding and execution-level precision, enabling more generalizable automated security testing frameworks.

\vspace{1mm}

\noindent \textbf{Limitations.}
While effective, \pq~is not without constraints:
\begin{itemize}[leftmargin=*, label=$\triangleright$]
    \item \textbf{Execution cost and rate limits.} LLM-guided fuzzing remains computationally intensive, and API rate throttling can slow feedback cycles.
    \item \textbf{Context window constraints.} Even with retrieval augmentation, large schemas can exceed model context limits, leading to partial prompt conditioning and missed relationships.
    \item \textbf{Response interpretation ambiguity.} The context-aware detector can identify likely vulnerabilities, but some cases still require human validation to confirm exploitability.
    \item \textbf{Model bias and non-determinism.} Variations in model architecture and decoding strategies lead to inconsistent results, motivating ensemble or calibration techniques for reproducibility.
\end{itemize}

\noindent \textbf{Future Work.} The growing ecosystem of agentic LLM frameworks offers a natural evolution path for \pq. A multi-agent design, with specialized agents for query generation, evaluation, and refinement, could enable continuous self-improvement and deeper exploit discovery. Another direction is the development of domain-specialized LLMs for API and schema reasoning, analogous to text-to-SQL models, which could reduce prompt overhead while improving precision and generalization. Finally, exploring hybrid systems that couple LLM reasoning with program analysis or formal verification could enable both semantic adaptability and provable assurance.

\section{Conclusion}
\pq~shows that the combination of retrieval, reasoning, and adaptive learning can fundamentally improve the way GraphQL APIs are tested. By integrating large language models into the fuzzing loop, it transforms random exploration into guided reasoning, allowing the system to understand schemas, infer dependencies, and generate purposeful queries. Through multi-armed bandit strategy selection, \pq~learns which testing behaviors yield the most valuable feedback, achieving higher coverage and uncovering vulnerabilities that existing tools consistently miss. Beyond its empirical gains, \pq~reveals a deeper insight: LLMs can act not only as generators but as analysts that interpret system behavior. This ability to connect input, responses, and context enables the detection of complex logic-level flaws that evade rule-based or pattern-driven scanners. These results mark a step towards autonomous, self-improving security testing, where models learn from every execution to test smarter over time. Future extensions will explore collaborative, multi-agent setups and large-scale retrieval across heterogeneous APIs, paving the way for intelligent systems that continuously learn the structure and weaknesses of modern Web applications.

\newpage
\medskip
\onecolumn \begin{multicols}{2}

\bibliography{arxiv}
\end{multicols}

\end{document}